\documentclass[conference,letterpaper]{IEEEtran}
\IEEEoverridecommandlockouts
\usepackage[T1]{fontenc}
\usepackage{cite}
\usepackage{booktabs}
\usepackage{multirow}
\usepackage{amsmath}
\usepackage{amssymb}
\usepackage{graphicx}
\usepackage[table]{xcolor}

\definecolor{blendgray}{gray}{0.93}

\AtBeginDocument{%
  }

\begin{document}

\pagestyle{empty}

\title{DeGS: A Scalable 3DGS Architecture via Decoupled Workload Parsing and Reorganization}

\author{%
\IEEEauthorblockN{Minnan Pei\textsuperscript{1,2},
Gang Li\textsuperscript{1,\textdagger},
Zeyu Zhu\textsuperscript{1,2},
Siting Wang\textsuperscript{1,2},
Junwen Si\textsuperscript{1},\\
Zhuoran Song\textsuperscript{3},
Yu Feng\textsuperscript{3},
Fangxin Liu\textsuperscript{3},
Xiaoyao Liang\textsuperscript{3}, and
Jian Cheng\textsuperscript{1,2,\textdagger}}
\IEEEauthorblockA{\textsuperscript{1}\textit{Institute of Automation, Chinese Academy of Sciences, C\textsuperscript{2}DL}, Beijing, China}
\IEEEauthorblockA{\textsuperscript{2}\textit{University of Chinese Academy of Sciences}, Beijing, China}
\IEEEauthorblockA{\textsuperscript{3}\textit{Shanghai Jiao Tong University}, Shanghai, China\\
peiminnan19@mails.ucas.ac.cn}
\thanks{\textsuperscript{\textdagger}Gang Li and Jian Cheng are the co-corresponding authors.}
}

\maketitle
\thispagestyle{empty}

\begin{abstract}
3D Gaussian Splatting (3DGS) has emerged as a leading technique for real-time novel view synthesis, yet existing 3DGS accelerators suffer from poor architectural scalability: increasing the number of PEs leads to marginal performance improvement during rendering. We identify that the root cause is the tightly coupled ``checking-while-blending'' dataflow, which exacerbates PE underutilization caused by spatial redundancy from irregular Gaussian coverage and temporal redundancy from asynchronous pixel-wise termination under parallel execution.

To address this issue, we propose DeGS, a scalable architecture for efficient 3DGS inference. To systematically eliminate the redundancies inherent in rendering, DeGS exploits a decoupled dataflow, restructuring the coupled $\alpha$-checking, transmittance checking, and $\alpha$-blending of the standard rendering process into consecutive workload parsing, reorganization, and blending stages. This allows the fragmented, length-variable, and temporal-dependent workloads to be reorganized into compact, conflict-free, and dense workloads prior to blending, thereby significantly improving PE utilization during parallel blending.
Implemented in 28 nm technology, DeGS achieves 2.36$\times$--7.25$\times$ throughput, 1.82$\times$--6.02$\times$ end-to-end speedup, and 1.59$\times$--4.42$\times$ energy efficiency over state-of-the-art 3DGS accelerators (GSCore, GBU, GCC) across diverse scenes and resolutions (720p to 8K). Moreover, scaling from 16 to 1024 PEs, DeGS maintains over 80\% PE utilization at high resolutions, significantly outperforming existing accelerators.
             
\end{abstract}


\section{Introduction}
\label{sec:intro}

3D Gaussian Splatting (3DGS) has rapidly emerged as a compelling technique for novel view synthesis, combining high visual quality and real-time rendering capabilities \cite{3dgs, survey1, survey2, survey3, survey4, 2dgs, 3dgszip, 4d, 4dgs, ges, gs-scale, adr}. It has demonstrated outstanding performance in applications such as AR/VR \cite{vr-gs, vr1,vr2,vr3,vr4}, SLAM-based navigation \cite{wildgs,cg, rtg, sgs, lic, splatam}, and physical environment simulation \cite{gaussianproperty,physgaussian,physics3d,ready,splatad}. In such diverse scenarios, 3DGS often needs to operate under varying input resolutions (ranging from 720p to 8K) as well as different computational and power budgets. While a number of efficient 3DGS accelerators have been proposed for specific applications \cite{metasapiens, gcc, gsnorm, gscore, gbu}, endowing them with strong architectural scalability remains critical.

According to the operation characteristics, the standard 3DGS processing pipeline can be broadly divided into two stages: memory-intensive Gaussian preprocessing and compute-intensive Gaussian rendering. In the preprocessing stage, the learned 3D Gaussian representation is converted into a 2D form through Gaussian culling, projection, and sorting. The rendering stage then performs pixel-wise $\alpha$-blending on these 2D Gaussians to synthesize the final RGB image. Although the first stage involves massive memory operations, its efficiency can be effectively improved by increasing DRAM bandwidth \cite{gcc}, which makes the second stage dominate the overall execution time \cite{flashgs, characterization, adr, speedy-splat}. 
For compute-intensive rendering, performance is expected to scale with available computational power.
However, we observe that existing 3DGS accelerators fail to exhibit such architectural scalability: \textit{increasing the number of compute units often leads to a significant drop in their utilization, resulting in marginal performance improvement.} For instance, in 2K image rendering, as the number of blending processing elements (PEs) is increased from 16 to 1024, the PE utilization of GSCore \cite{gscore} and GBU \cite{gbu} drastically drops from 84.7\% and 18.7\% to 34.8\% and 7.1\%, respectively, resulting in a performance improvement of only 6.28$\times$ and 8.47$\times$. We identify that the primary cause of this gap lies in the tightly coupled rendering dataflow adopted in existing accelerators, which exacerbates PE underutilization caused by spatial and temporal redundancies in parallel processing. This scalability gap stems from two factors:


\textbf{The mismatch between irregular, dynamic workloads and regular parallel processing results in massive redundancy.} The 3DGS representation exhibits an inherent irregular and dynamic nature. In the 2D spatial dimension, the Gaussian splats vary in shape and size. Existing coarse-grained tile-based (e.g., 16$\times$16) parallel rendering computes a massive number of pixels outside the effective Gaussian regions, which do not contribute to the final image, resulting in \textit{spatial redundancy}. In the temporal dimension, each pixel's blending proceeds in a near-to-far depth order, and the termination time of this process differs across pixels. When blending is performed on a per-tile basis, PEs that finish early cannot proceed to new tasks until all pixels within the same tile complete their blending, leading to \textit{temporal redundancy}. As the tile size and number of PEs increase, both forms of redundancy are significantly exacerbated, causing a substantial decline in PE utilization.

\textbf{The tightly coupled condition checking and pixel blending dataflow hinders the effective elimination of aforementioned redundancy.} 
While spatial and temporal redundancies in the rendering process are readily identifiable, existing accelerators still struggle to eliminate such redundant computations. The fundamental reason lies in their adherence to the standard blending pipeline, where pixel validity checking, pixel termination condition checking, and pixel blending are executed simultaneously in each blending iteration due to the shared $\alpha$ calculation. This prevents a pixel from being predetermined as valid before $\alpha$ calculation is completed, introducing massive ineffective computations caused by spatial and temporal redundancy. In summary, the tightly coupled ``checking-while-blending'' rendering dataflow makes it difficult to identify redundancies in advance and dynamically skip them for higher effective PE utilization.

In this paper, we present DeGS, a scalable architecture for efficient 3DGS inference. Unlike the ``checking-while-blending'' rendering dataflow in existing works, our core idea is to decouple condition checking from pixel blending, enabling blending to utilize computing resources more efficiently during computational scaling. Specifically, we reformulate the rendering process as three consecutive stages: workload parsing, workload reorganization, and blending. This is achieved by an efficient span analysis that can identify the exact scope of effective workloads, avoiding on-the-fly pixel-wise validation. Prior to blending, we explicitly parse the fragmented, length-variable, and temporal-dependent workloads, extract fine-grained valid workload segments, and reorganize them into compact, conflict-free, and dense ones. In this way, we can eliminate spatial and temporal redundancies as much as possible and improve effective PE utilization during rendering. 
This paper makes the following contributions:
\begin{itemize}
\item \textbf{Decoupled Dataflow:} We restructure the standard rendering pipeline so that redundant work is removed before color blending. Span Parsing uses 1D scanline solving to identify valid spans before pixel-wise $\alpha$ computation, while Task Reorganization then groups spans with varying lengths and dependencies into packets that can be executed without conflicts. Blending Execution finally processes only these regularized packets, enabling dense and scalable rendering. Together, these stages move validity checking and dependence handling out of the blending array, allowing the backend to execute only regularized valid work.
    
\item \textbf{Scalable Architecture:} We propose DeGS, a 3DGS rendering architecture that isolates irregular workload handling from backend blending. Span Engine performs front-end row-span parsing in hardware, removing coarse boundary processing from the backend. Packing Scheduler then converts these variable-length, dependence-constrained spans into dependence-safe fixed-width packets through banked distribution, conflict-aware issuing, and packet compaction. As a result, the Blending Array operates on a regular packet interface and can be specialized purely for arithmetic throughput, rather than on-the-fly validity checking or hazard resolution.


\item \textbf{State-of-the-art Performance:} We implement our accelerator in a 28nm technology node. Evaluation shows that DeGS achieves 2.36$\times$--7.25$\times$ throughput, 1.82$\times$--6.02$\times$ end-to-end speedup, and 1.59$\times$--4.42$\times$ energy efficiency over state-of-the-art 3DGS accelerators across scenes. Moreover, across the evaluated design scales from 16 to 1024 PEs and resolutions from 720p to 8K, DeGS consistently provides the best PE utilization, latency, throughput, and energy efficiency.

\end{itemize}

\section{Background and Motivation}
\label{sec:background}

\begin{figure*}[t]
  \centering
  \includegraphics[width=\textwidth]{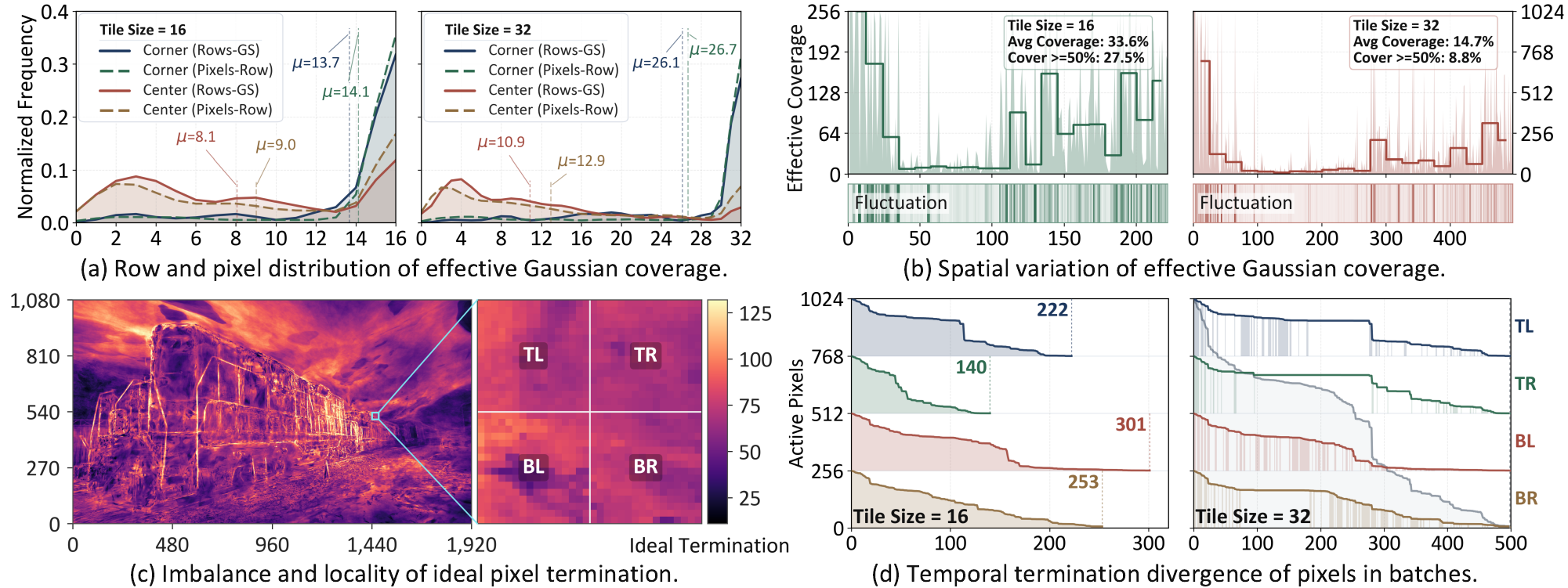}
  \caption{Analysis of intrinsic spatial and temporal workload irregularity in Gaussian blending.}
  \label{fig:fig1}
\end{figure*}

\subsection{Standard 3DGS Processing Pipeline}
\label{sec:bg_pipeline}

3D Gaussian Splatting (3DGS) represents a scene as a collection of 3D spatial Gaussian attributes and a standard 3DGS rendering pipeline can be divided into two stages: \emph{Preprocessing}, which transforms each 3D Gaussian into a 2D screen-space footprint and derives its screen-space parameters in the depth and tile order; \emph{Rendering}, which evaluates pixel-wise Gaussian contributions and performs $\alpha$-blending to generate the final image. 

For the $i$-th 3D Gaussian, its geometry is parameterized by a mean vector $\mu_i$ and a covariance matrix $\Sigma_i$, while its view-dependent appearance is represented by spherical harmonics coefficients $\mathrm{sh}_i$. Given the view transformation matrix $W$ and the projection function $\pi(\cdot)$, the projected screen-space center and the corresponding view-dependent color are given by:
\begin{equation}
\mu_i^{*} = \pi(W\mu_i), \qquad c_i = f(W; \mu_i; \mathrm{sh}_i),
\label{eq:center_color}
\end{equation}
where $f(\cdot)$ denotes the spherical harmonics evaluation function. 

Let $J$ denote the Jacobian of the projection function at the transformed Gaussian center. Using the standard affine approximation, the corresponding 2D screen-space covariance matrix is:
\begin{equation}
\Sigma_i^{*} = J \left( W \Sigma_i W^\top \right) J^\top.
\label{eq:proj_cov}
\end{equation}
Its inverse covariance is parameterized as:
\begin{equation}
(\Sigma_i^{*})^{-1} =
\begin{bmatrix}
A & B \\
B & C
\end{bmatrix},
\label{eq:3}
\end{equation}
where $(A,B,C)$ compactly describe the projected Gaussian shape in screen space. In practical implementations, these projected parameters are further used to derive a coarse 3-$\sigma$ bounding box. Each Gaussian is paired with the tiles on this bounding box, and then sorted in near-to-far depth order for each tile. 

For a screen-space sample at pixel coordinate $p=(x,y)$, let:
\begin{equation}
dx = x-\mu_{i,x}^{*}, \qquad dy = y-\mu_{i,y}^{*}.
\label{eq:4}
\end{equation}
The corresponding quadratic form is:
\begin{equation}
\begin{aligned}
q(x,y)
&=
\begin{bmatrix}
dx & dy
\end{bmatrix}
\begin{bmatrix}
A & B \\
B & C
\end{bmatrix}
\begin{bmatrix}
dx \\
dy
\end{bmatrix}
\\
&=
A\,dx^2 + 2B\,dx\,dy + C\,dy^2.
\end{aligned}
\label{eq:5}
\end{equation}
This quadratic form determines both the effective support of the Gaussian in screen space and its opacity contribution at each sample location.

\begin{figure*}[t]
  \centering
  \includegraphics[width=\textwidth]{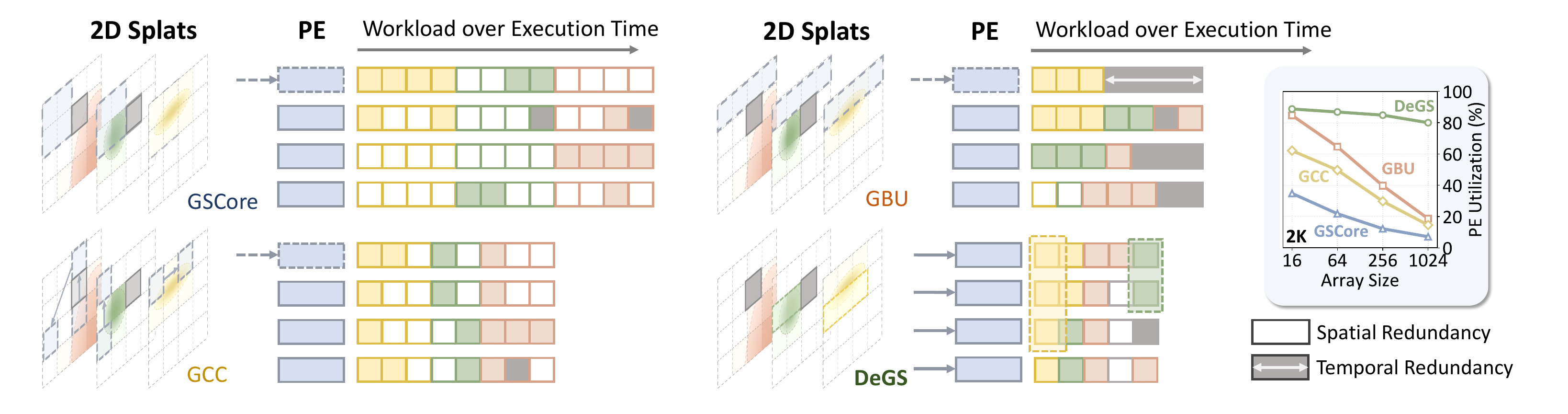}
  \caption{Qualitative comparison of workload regularity and PE utilization across 3DGS accelerators. DeGS reduces both spatially invalid work and temporal waiting.}
  \label{fig:fig2}
\end{figure*}

In the blending stage, each projected Gaussian is sampled on the pixel grid, producing a set of candidate \emph{fragments}. For a fragment associated with pixel $p$, the corresponding opacity is:
\begin{equation}
\alpha_p = o_i \cdot \exp\!\left(-\frac{1}{2} q(x,y)\right),
\label{eq:6}
\end{equation}
where $o_i$ is the base opacity of the Gaussian primitive. Since a Gaussian has infinite mathematical support, standard 3DGS introduces a strict truncation threshold $\epsilon$ (typically $1/255$) and performs $\alpha$-checking to determine whether a candidate fragment contributes non-negligible opacity. Specifically, a fragment is considered valid only if $\alpha_p \ge \epsilon$, which is equivalently expressed as the following analytic ellipse in screen space:
\begin{equation}
q(x,y) \le Q_{\text{th},i} \quad \text{where} \quad Q_{\text{th},i} = -2\ln(\epsilon / o_i).
\label{eq:7}
\end{equation}
Fragments outside this boundary are discarded and do not participate in the subsequent blending process.

For fragments that pass $\alpha$-checking, 3DGS performs standard near-to-far $\alpha$-blending. Let $C_p$ denote the accumulated color of pixel $p$, and let $T_{p,i}$ denote the residual transmittance before processing the $i$-th Gaussian. The blending recurrence is:
\begin{equation}
C_p \leftarrow C_p + T_{p,i}\,\alpha_p\,c_i, \qquad
T_{p,i+1} \leftarrow T_{p,i}(1-\alpha_p),
\label{eq:8}
\end{equation}
where $c_i$ is the view-dependent color of the Gaussian. Since $T$ decreases monotonically along the recurrence, standard 3DGS also employs \emph{asynchronous early termination}: once $T_{p,i} < \tau$ for a small threshold $\tau$ (e.g., $10^{-4}$), the pixel is considered converged and all remaining farther Gaussians are skipped for that pixel.

\textcolor{black}{
At a high level, 3DGS rendering projects each learned 3D Gaussian into a 2D elliptical footprint, generates candidate fragments from a coarse screen-space bounding box, and blends only the fragments that pass the opacity threshold. For a projected Gaussian overlapping a $16\times16$ tile, the bounding box may cover many candidate pixels, while the true valid support is only the elliptical region satisfying Eq.~\ref{eq:7}. Valid fragments are then processed in near-to-far order using the transmittance recurrence in Eq.~\ref{eq:8}, and different pixels may terminate at different depths once their residual transmittance becomes small. This first-principles example directly exposes the two sources of ineffective parallel work in standard tile-based rendering: invalid candidates outside the Gaussian support cause spatial redundancy, and early-terminated pixels waiting for long-tail pixels cause temporal redundancy.
}

\subsection{Scalability Challenges of 3DGS Architecture}
\label{sec:limitations}

Based on the operation characteristics, the 3DGS processing pipeline can be broadly categorized into two stages: a memory-intensive preprocessing stage, which includes Gaussian culling, projection, and sorting, and a compute-intensive rendering stage, comprising $\alpha$ computation, termination condition checking, and pixel blending. As for preprocessing, although it involves extensive memory operations, existing work has shown that its efficiency can be effectively improved by increasing DRAM bandwidth (e.g., by replacing LPDDR4 with LPDDR5) \cite{gcc,neo}. This makes the rendering stage the primary performance bottleneck. 

To enhance rendering efficiency, a series of dedicated 3DGS accelerators have been proposed \cite{gscore,gbu,gcc}. However, we observe a clear scalability issue: \textit{as the number of PEs increases, their utilization drops significantly, leading to marginal gains in rendering performance}. Through in-depth analysis, we identify that the underlying cause lies in the coupled dataflow employed by existing architectures, which prevents the effective elimination of redundancies during rendering.

\subsubsection{Spatial and Temporal Redundancy}
The 2D Gaussians obtained by preprocessing exhibit significant irregularity, and this irregularity demonstrates evident dynamic variations throughout the rendering process. In the 2D spatial dimension, Gaussians at a fixed viewpoint vary in shape and size, leading to drastically different projection coverages: some fine-grained Gaussians cover only a few pixels, whereas others may span tens of pixel rows or even larger regions, as illustrated in Figure~\ref{fig:fig1}(a). While reducing the tile size can locally decrease the number of invalid pixels, it increases the number of tile intersections per Gaussian and thus introduces a much heavier bandwidth bottleneck~\cite{gcc, neo}.
Moreover, as shown in Figure~\ref{fig:fig1}(b), the number of valid pixels covered by a specific tile also varies significantly during rendering along the depth direction. In tile-based rendering, computations on different pixels within a tile are executed in parallel. This leads to a large number of invalid pixels outside Gaussians occupying compute units during rendering. We refer to this type of redundancy as \textbf{spatial redundancy}.

In the depth direction, the blending process for each pixel employs a strict near-to-far iterative computation scheme, terminating when the residual transmittance falls below a threshold. During this process, the execution depths of different pixels exhibit both locality and notable variation. As shown in Figure~\ref{fig:fig1}(c), adjacent pixels often terminate computation in similar batches, whereas pixels that are farther apart exhibit substantial differences in termination time. 
This results in a significant imbalance in the completion times of pixels within a fixed-size tile. During parallel execution, computations that finish early are forced to wait for the remaining ones to complete because of Eq.~\ref{eq:8}, leading to a waste of computational resources. Meanwhile, as tile size increases under higher degrees of parallelism, pixel blending that could otherwise terminate early is increasingly delayed by long-tail pixels within the batch, as Figure~\ref{fig:fig1}(d) shows. We refer to this idling of computing resources in parallel processing, caused by varying execution depths, as \textbf{temporal redundancy}.

In conclusion, the mismatch between irregular, dynamic workloads and coarse-grained regular parallel processing results in massive redundancy during rendering.

\begin{figure*}[t]
  \centering
  \includegraphics[width=\textwidth]{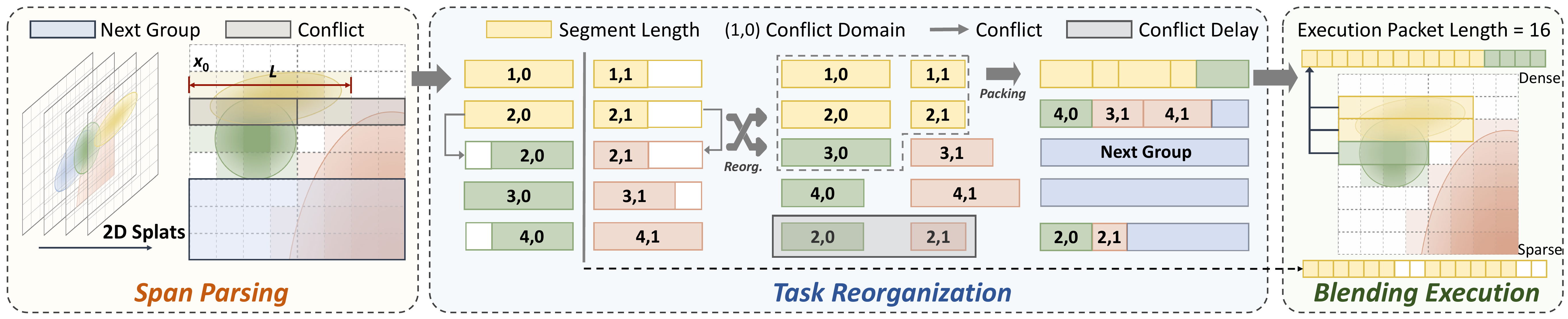}
  \caption{\textcolor{black}{Decoupled DeGS dataflow. Span Parsing extracts valid row spans from projected Gaussians, Task Reorganization converts length-variable dependent spans into conflict-free packets, and Blending Execution operates on regularized packets.}}
  \label{fig:fig3}
\end{figure*}

\subsubsection{Coupled Condition Checking and Pixel Blending}
\label{sec:2.2.2}
Although spatial and temporal redundancies can be identified, the tightly coupled dataflow of condition checking and blending adopted by existing architectures makes it difficult to avoid the resulting performance penalty. Concretely, the rendering of a given tile comprises three components: pixel validity checking ($\alpha$-checking), pixel blending ($\alpha$-blending), and pixel termination condition checking (transmittance checking). Since $\alpha$ computation is involved in all three steps, the pipeline effectively forms a coupled, ``checking-while-blending'' mechanism. That is, the checking result at time $T$ can only guide the computation at time $T+1$, while the invalid computations at time $T$ itself cannot be eliminated. This results in redundancy in every iteration of the blending process. As the number of PEs and the degree of computational parallelism increase, both spatial and temporal redundancies are exacerbated. This coupled execution pattern further reduces the PE utilization, making it difficult for blending performance to scale effectively with the increased computational power. Throughout this paper, unless otherwise specified, \textit{PE utilization} refers exclusively to the utilization of the pixel-wise calculation in Eq.~\ref{eq:6} and Eq.~\ref{eq:8}. Any slot spent on extra computation and idle cycles caused by spatial redundancy or temporal redundancy is counted as utilization loss.

\subsubsection{Limitation of Existing 3DGS Architectures}

Figure~\ref{fig:fig2} shows three typical "checking-while-blending" coupled dataflows in prior 3DGS accelerators. \textbf{GSCore} \cite{gscore} adopts the most rigid regular dataflow: it processes a fixed 2D Gaussian splat region every cycle regardless of the valid coverage or asynchronous termination. As in the standard coupled rendering flow, invalid fragments occupy PE slots in the spatial dimension, while lockstep execution forces early-terminated pixels to wait for the long-tail pixels in the temporal dimension. As a result, GSCore suffers from both spatial redundancy and temporal redundancy most directly, and therefore achieves the lowest PE utilization. 
\textbf{GBU} \cite{gbu} and \textbf{GCC} \cite{gcc} both generate the subsequent workload on the fly by coupling $\alpha$-checking with boundary testing at each execution step, thereby reducing part of the spatial redundancy. 
In GBU, each row is bound to a specific PE because of the read-modify-write dependence in Eq.~\ref{eq:8}. Consequently, irregular workloads across rows cannot be redistributed after assignment. Splat irregularity appears as unequal waiting windows across rows spatially, while temporal irregularity appears as idle PE lanes after early termination, and both become more severe as the array scales up. 
GCC, however, still relies on a fixed-size traversal window mapped onto regular pixels inside a coupled evaluate-while-checking flow. As the array grows, the search window grows as well, making the filtering increasingly coarse and gradually reintroducing GSCore-like spatial and temporal redundancy. Overall, these architectures differ only in how irregular work enters execution, but none of them decouples redundancy identification from blending execution before work reaches the PE array.

\subsection{Our Approach}

To achieve notable performance scaling with an increasing number of PEs for time-dominant rendering, the key lies in reducing the spatial and temporal redundancies mentioned earlier. However, due to the coupled rendering dataflow that intertwines condition checking and pixel blending, existing accelerators suffer from severe computational waste during parallel processing.

To boost effective PE utilization during rendering, an intuitive approach is to eliminate redundancy by identifying and reorganizing the fragmented and sparsely distributed workloads into compact and dense ones for subsequent blending. Accordingly, we present a decoupled rendering dataflow, which exploits an efficient span analysis method to locate valid rendering workloads in advance of blending, requiring neither on-the-fly checking nor brute-force pre-computation of all $\alpha$ values. In this way, the standard rendering process is reformulated as consecutive workload parsing, reorganization, and blending stages, which helps to improve architectural scalability. 

\section{DeGS Dataflow}
\label{sec:dataflow}

DeGS decouples standard 3DGS blending into three concrete stages. First, Span Parsing directly extracts valid row spans from projected Gaussians, using lightweight computation to resolve the span region before execution. Second, Task Reorganization reschedules and packs dependence-constrained span tasks into conflict-free rendering packets. Finally, Blending Execution consumes only these regular packets and performs dense $\alpha$-blending. Importantly, this three-stage decoupling preserves the valid blending recurrence in Eq.~\ref{eq:8} while removing and absorbing workload irregularity before blending execution.

\subsection{Span Parsing}
\label{sec:span_parsing}

In the conventional coupled dataflow, hardware must evaluate Eq.~\ref{eq:6} for all candidate fragments inside a coarse 2D bounding box, and then use the condition $\alpha_p \ge \epsilon$ to determine whether each candidate actually falls inside the valid Gaussian support. Under this coupled checking-and-execution flow, however, fragments that are eventually discarded have already consumed issue bandwidth and incurred useless computation on both $\alpha_p$ and Eq.~\ref{eq:8} in the current iteration. As a result, the large number of invalid candidate fragments introduced by irregular spatial coverage directly appears as spatial redundancy in the array.

Rather than filtering invalid fragments inside the coupled blending computation, \emph{Span Parsing} directly extracts valid row spans before execution. 
By exploiting the analytic ellipse of projected 2D Gaussians, it transforms per-fragment probing in a coarse 2D bounding box into a 1D scanline problem without explicitly evaluating $\alpha_p$ at every pixel. Instead, it leverages the analytic continuity of the ellipse to directly recover the length of each valid interval.

Specifically, the valid support of a 2D Gaussian is defined by Eq.~\ref{eq:7}. Since this iso-contour is an analytic ellipse, its intersection with any fixed scanline $y$ must be a contiguous 1D interval. Therefore, if we can efficiently determine the span center and half-width on each scanline, we can directly recover the full valid row span without probing all pixels inside the bounding box. A direct row solution of $q(x,y)$ in Eq.~\ref{eq:5} for $x$ is mathematically valid, but introduces expensive algebraic operations into the rasterization path, offsetting the benefit of exact boundary trimming. To avoid this cost, we first rewrite the original quadratic form as:
\begin{equation}
q(x,y)
=
A\left(dx+\frac{B}{A}dy\right)^2
+
\left(C-\frac{B^2}{A}\right)dy^2,
\label{eq:q_reform}
\end{equation}
and extract a set of structural parameters that depend only on the intrinsic Gaussian geometry:
\begin{equation}
m=-\frac{B}{A}, \qquad
C' = C-\frac{B^2}{A},
\label{eq:m_cprime}
\end{equation}
\begin{equation}
r_x=\sqrt{\frac{Q_{\text{th}}}{A}}, \qquad
\mathrm{inv\_}r_y^2=\frac{C'}{Q_{\text{th}}}.
\label{eq:rx_ry}
\end{equation}

These parameters are mathematically equivalent to the original covariance parameters $(A,B,C,Q_{\text{th}})$, but are much better suited for direct scanline boundary solving. More importantly, they can be precomputed during the memory-dominated preprocessing stage and passed into the rendering pipeline as compact reusable descriptors.

With these parameters, for a given scanline offset $dy$, the center of the valid row span is:
\begin{equation}
x_c(y)=\mu_x+m\cdot dy,
\label{eq:span_center}
\end{equation}
and substituting the transformed parameters into the support condition:
\begin{equation}
A(x-x_c)^2 + C'dy^2 \le Q_{\text{th}}
\label{eq:span_constraint}
\end{equation}
gives the half-width of the valid span:
\begin{equation}
w(dy)=r_x\cdot \sqrt{1-dy^2\cdot \mathrm{inv\_}r_y^2}.
\label{eq:span_halfwidth}
\end{equation}
\begin{equation}
\begin{aligned}
x_0&=\left\lfloor x_c-w \right\rfloor-\delta, &
x_1&=\left\lceil x_c+w \right\rceil+\delta,\\
L&=x_1-x_0+1.
\end{aligned}
\end{equation}

With \(w\) and \(x_c\), we can directly determine the starting position \(x_0\) and valid length \(L\) of each row span, while \(\delta\) estimates the error based on $\frac {dw} {dy}$. However, if we continue evaluating $\alpha_p$ in Eq.\ref{eq:6}, the extra cost of span solving would lead to higher overall complexity and area overhead. To offset this cost and further exploit the continuity of the 2D Gaussian, we must compactly encode $q(x,y)$ within the span. Let the screen coordinate be discretized as $x=x_0+n$, and define the row sequence:
\begin{equation}
g_n = q(x_0+n, y),
\label{eq:gn}
\end{equation}
with first-order forward difference:
\begin{equation}
d_n = g_{n+1}-g_n.
\label{eq:dn}
\end{equation}
Classical digital differential analyzer (DDA) rasterization yields:
\begin{equation}
g_{n+1}=g_n+d_n, \qquad d_{n+1}=d_n+2A.
\label{eq:dda}
\end{equation}
However, this pixel-wise recurrence introduces strict loop-carried dependence and would reintroduce serialization into blending. To avoid that, we derive the closed form:
\begin{equation}
q_n = g_0 + n\cdot d_0 + A\cdot n(n-1).
\label{eq:18}
\end{equation}
As a result, one complete valid \emph{row span} can be compactly represented as:
\begin{equation}
(y, x_0, L, g_0, d_0, A),
\label{eq:span_packet}
\end{equation}
while $(g_0,d_0)$ preserve the minimum state required to reconstruct the quadratic value of any valid fragment within the span. Blending execution no longer needs to perform exploratory boundary probing, instead, it only reconstructs $q_n$ from the local offset $n$.

In summary, Span Parsing resolves exact Gaussian row spans with lightweight row-level computation, while reducing pixel-wise evaluation from Eq.~\ref{eq:5} to the closed-form reconstruction in Eq.~\ref{eq:18}. The key effect is not merely earlier filtering, but no longer feed the backend execution with a coarse candidate region, but a compact row span as a small set of parameters that contains only valid state required for reconstructing complete information.

\begin{figure*}[t]
  \centering
  \includegraphics[width=\textwidth]{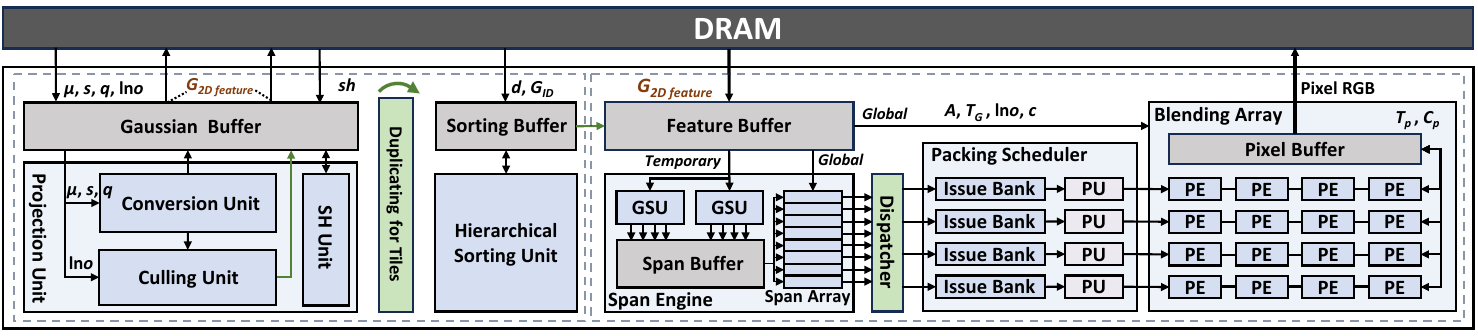}
  \caption{Overall design of DeGS architecture.}
  \label{fig:fig4}
\end{figure*}

\subsection{Task Reorganization}
\label{sec:task_reorg}

After Span Parsing, invalid fragments have been removed, but the remaining row spans are still irregular in length and constrained by pixel-wise update order. \emph{Task Reorganization} therefore makes the dependence structure explicit. We partition spans into \emph{conflict domains}, where segments in the same domain may update overlapping pixel state and therefore must preserve depth order, while segments from different domains can be issued opportunistically. The scheduler then converts variable-length spans into fixed-width packets only after ordering and reuse-distance constraints have been enforced.

In coupled dataflow, if multiple tasks update the same pixel state at nearby times, directly feeding these spans into blending in generation order would still allow local dependence conflicts to reappear as long-tail stalls and execution bubbles.
For decoupling, dependence-aware scheduling and packing is performed. A conflict domain is the smallest local scheduling region in this reorganization. Across different conflict domains, segments can be issued out of order whenever resources are available to maximize parallelism. Within the same domain, however, issuance must preserve the original depth order and satisfy a minimum conflict delay to avoid read-modify-write hazards. After dependence-safe issuance, the resulting segments are assembled into fixed-width \emph{task packets}, where each row span is decomposed into pixel-level tasks. Conflict-free tasks drawn from different spans are then packed together into a fixed-width rendering packet. As a result, temporal imbalance is confined to bounded scheduling regions rather than amplified at the array level, and the backend receives a continuous, regularized task stream for dense execution.

As illustrated in Figure~\ref{fig:fig3}, when a later task $(2,1)$ conflicts with the just-issued task $(2,0)$ in the same row, it is temporarily stalled due to the four-cycle conflict delay, and the scheduler instead steals a ready task such as $(3,0)$ or $(4,0)$ from an idle domain. This out-of-order peeking mechanism fundamentally breaks the conventional tile-level lockstep execution model: long-tail rows now cause only very localized blocking, and the scheduler can immediately interleave short tasks from other domains. As a result, early-converged regions and long-tail regions are naturally overlapped on the timeline, eliminating global temporal idling at the tile level.

\subsection{Blending Execution}
\label{sec:blending_exec}
After Span Parsing and Task Reorganization, the backend no longer receives raw candidate fragments. Instead, it receives fixed-width packets whose validity has already been resolved and whose update order has already been made dependence-safe. \emph{Blending Execution} therefore performs only three functions: reconstruct $q_n$ in Eq.~\ref{eq:18} from the span descriptor, evaluate $\alpha$, and apply the standard recurrence. For each fragment in execution packet of valid row spans, Blending Execution computes:
\begin{equation}
\alpha = \exp\!\left(-\frac{1}{2}q_n\right),
\label{eq:6_backend}
\end{equation}
and execute the transmittance and color recurrence in Eq.~\ref{eq:8}.

In this way, the coupled ``checking-while-blending'' execution flow is fully broken: spatial redundancy has been removed by Span Parsing, temporal redundancy has been absorbed by Task Reorganization, and Blending Execution operates only on regularized valid work. As a result, redundancy is no longer exposed to the blending process, but is instead eliminated or absorbed before arithmetic execution while preserving the standard 3DGS blending semantics.

\section{DeGS Microarchitecture}
\label{sec:microarchitecture}

As motivated in Section~\ref{sec:background}, the scaling bottleneck of prior 3DGS accelerators is not blending arithmetic itself, but the direct exposure of irregular Gaussian coverage and asynchronous pixel termination to a wide PE array. DeGS is therefore organized around irregularity isolation.
Specifically, the Span Engine prevents spatially irregular and partially invalid Gaussian support from entering the backend as coarse candidate fragments by converting each projected Gaussian into valid row spans. The Packing Scheduler then prevents variable span length and update-order constraints from propagating to the backend by issuing only dependence-safe, fixed-width packets. As a result, the Blending Array sees a regular packet stream rather than raw irregular rendering work, and can be optimized purely for arithmetic throughput.
By isolating spatial invalidity and temporal dependence before arithmetic execution, DeGS prevents frontend irregularity from propagating into the Blending Array, which is the key to scalable PE utilization.

\subsection{Span Engine}
\label{sec:span_engine}

The \emph{Span Engine} converts each projected Gaussian into a stream of row spans before the Gaussian reaches the Packing Scheduler. It enumerates only the scanlines that intersect the Gaussian valid coverage and emits one Span Vector for each valid scanline. By moving this work out of the Blending Array, DeGS prevents coarse 2D bounding boxes from expanding into invalid pixel tasks in the arithmetic pipeline.

\begin{figure}[t]
  \centering
  \includegraphics[width=0.48\textwidth]{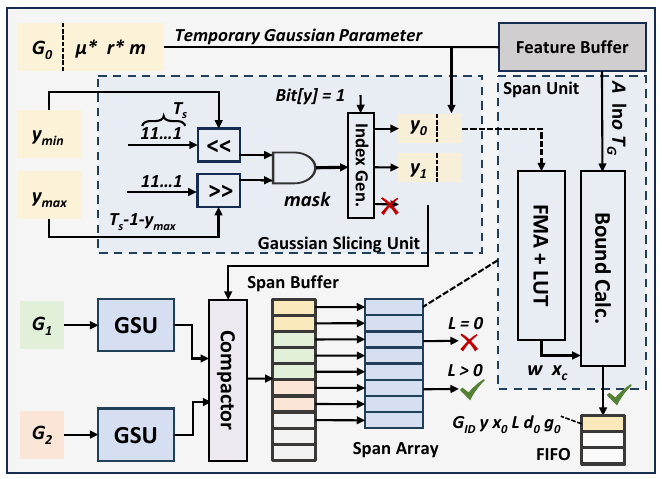}
  \caption{Architecture of Span Engine.}
  \label{fig:fig5}
\end{figure}

As shown in Figure~\ref{fig:fig5}, the Span Engine contains two compute blocks with different throughput requirements. The Gaussian Slicing Unit (GSU) performs scanline enumeration. Using the temporary geometric descriptor produced during preprocessing, the GSU generates the valid scanline index range $[y_{\min}, y_{\max}]$ with shift-and-mask logic and pushes the resulting row work items into the Span Buffer. The Span Unit then consumes these row work items and solves the exact horizontal support of each scanline. For scanline $y$, the Span Unit computes the span center $x_c$, the span half-width $w$, and the initial closed-form parameters $(g_0,d_0)$ required by Eq.~\ref{eq:18}, and packs them into a \emph{Span Vector} $(\mathrm{G_{ID}}, y,x_0,L,g_0,d_0)$.

The Span Buffer is required because scanline enumeration and exact span solving have very different costs. The GSU is lightweight and should run continuously, whereas the Span Unit contains the critical arithmetic for span solving. In particular, the Span Unit evaluates the non-linear term in Eq.~\ref{eq:span_halfwidth} through a small LUT specialized for $\sqrt{1-x}$ over the bounded input range $x\in[0,1]$, avoiding a general floating-point square-root unit. If the GSU were directly coupled to the Span Unit, the number of Span Units would have to scale with the peak row-generation rate, which would make Span Engine disproportionately expensive. The Span Buffer breaks the coupling row slicing and row spanning, while the number of Span Units is chosen only to match the demand of the Packing Scheduler. 

This decoupling also changes how Gaussian parameters are stored on chip. DeGS does not keep one unified Gaussian-parameter pool for the entire rendering flow. Instead, it splits the projected Gaussian descriptor into two parts according to lifetime. The \emph{Temporary Gaussian Parameter} contains only the short-lifetime geometric fields required by the GSU and the Span Unit, and is released after span generation. The \emph{Global Gaussian Parameter} contains the attributes still needed after span generation, such as the coefficients and appearance fields later consumed by the Blending Array. This split reduces unnecessary parameter residency in the early pipeline stages and avoids transporting large Gaussian records through the Span Engine.

\subsection{Packing Scheduler}
\label{sec:packing_scheduler}

The \emph{Packing Scheduler} prevents variable span length and workload dependence from reaching the Blending Array. Its input is the stream of row spans produced by the Span Engine, and its output is a stream of fixed-width pixel task packets. To perform this conversion, the Packing Scheduler adds three hardware functions that are absent from a directly coupled design: banked task distribution, dependence-safe segment issuing, and fixed-width packet assembly.

\begin{figure}[t]
  \centering
  \includegraphics[width=0.48\textwidth]{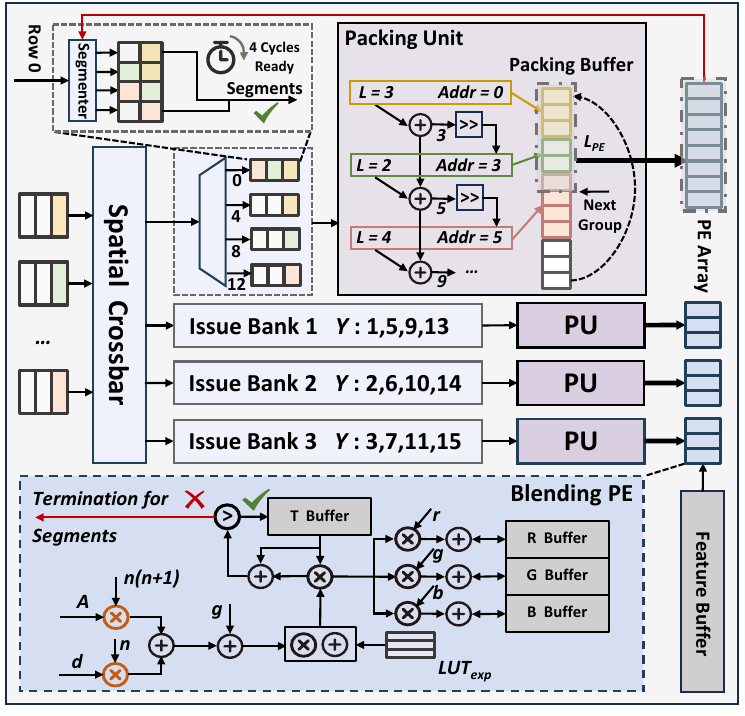}
  \caption{Architecture of Packing Scheduler \& Blending PE.}
  \label{fig:fig6}
\end{figure}

As shown in Figure~\ref{fig:fig6}, the Packing Scheduler first distributes incoming row spans across multiple Issue Banks through the Spatial Crossbar. The crossbar maps spans with different $y$ indices to different banks so that row-localized work is spread across the scheduling channels instead of accumulating in a single queue. After bank assignment, the \emph{Segmenter} splits each row span into fixed-stride Segments. A Segment is the minimum scheduling unit in the Packing Scheduler and the minimum spatial unit tracked for dependence control.

Dependence control is then applied at the Segment granularity. The first mechanism is termination feedback from the Blending Array. When the Blending Array detects that the pixels covered by a previously issued Segment have already satisfied the transmittance termination condition, it returns a termination signal to the corresponding Issue Bank, and the remaining farther Segments for that spatial region are discarded before issue. The second mechanism is a scoreboard-based delay constraint, which uses a timing wheel to detect overlapping updates to the same pixel region and enforces a fixed 4-cycle reuse distance, matching the write-back latency of the Blending Array. As a result, overlapping Segments are never issued within the unsafe read-modify-write interval.

After this dependence filtering, each Issue Bank forwards only issuable Segments to the Packing Unit (PU). Because the effective lengths of these Segments are still variable, the PU uses prefix-sum and shift logic to compact them into fixed-width pixel task packets. The resulting packets are stored in the Packing Buffer and then streamed to the Blending Array in order. Therefore, the Blending Array no longer receives raw row spans with uneven length and unresolved dependence. Instead, it receives only dense pixel packets whose ordering and reuse distance have already been enforced by the Packing Scheduler.

\subsection{Blending Array}
\label{sec:blending_array}

The Blending Array executes the pixel task packets produced by the Packing Scheduler. Its input is the fixed-width packet stream stored in the Packing Buffer, and its output is the updated pixel state in the Pixel Buffer. Because dependence ordering has already been enforced by the Packing Scheduler, the Blending Array only performs arithmetic execution and write-back.

In each cycle, each PE reads one pixel task from the current packet and extracts the local state required for closed-form span, including $(x,y)$, $(g_0,d_0)$, and the Gaussian identifier $\mathrm{G_{ID}}$. The packet does not carry the full Gaussian record. Instead, the PE uses $\mathrm{G_{ID}}$ to fetch the remaining Gaussian attributes from the Global Gaussian Parameter buffer, including the quadratic coefficient $A$ and the appearance parameters needed for blending. This late-fetch organization keeps the packet format compact and avoids carrying high-dimensional Gaussian attributes across the scheduling pipeline.

After operand collection, the PE reconstructs the exact quadratic value $q_n$ using Eq.~\ref{eq:18} and then computes the corresponding opacity $\alpha$. Since the local offset $n$ takes only a small set of discrete values inside each packetized segment, DeGS replaces a general-purpose multiplier with a specialized half-lookup FP-INT multiplier optimized for these offset cases. The resulting $q_n$ is then sent to LUT$_{\text{exp}}$ for opacity evaluation, after which the PE updates transmittance and accumulated color according to Eq.~~\ref{eq:8} and writes the updated pixel state back to the Pixel Buffer. Therefore, the Blending Array is a narrow and regular arithmetic stage: it consumes fixed-width packets, reconstructs $q_n$, evaluates $\alpha$, and performs the standard 3DGS blending recurrence without additional boundary testing or dependence arbitration.

\section{Evaluation}

\begin{table}[t]
\centering
\caption{Area and power breakdown of baselines and DeGS.}
\label{tab:t1}
\setlength{\tabcolsep}{3pt}
\renewcommand{\arraystretch}{1.05}

\begin{minipage}[t]{0.48\columnwidth}
\centering
\textbf{(a) Baselines}\par\smallskip
\resizebox{\linewidth}{!}{%
\begin{tabular}{l l c c c}
\toprule
\multicolumn{2}{c}{Architecture} & Area & Power & Config \\
\midrule
\multirow{4}{*}{GSCore}
 & P \& S     & 0.89 & 0.58 & -- \\
 & \cellcolor{blendgray}Blending
   & \cellcolor{blendgray}1.81
   & \cellcolor{blendgray}0.25
   & \cellcolor{blendgray}64 PE \\
 & \cellcolor{blendgray}SRAM
   & \cellcolor{blendgray}1.25
   & \cellcolor{blendgray}0.04
   & \cellcolor{blendgray}272 KB \\
\cmidrule(lr){2-5}
\vspace{0.25ex}
 & Total      & 3.95 & 0.87 & -- \\
\midrule
\multirow{5}{*}{GBU}
 & P \& S     & 0.89 & 0.58 & -- \\
 & \cellcolor{blendgray}Blending
   & \cellcolor{blendgray}2.88
   & \cellcolor{blendgray}0.88
   & \cellcolor{blendgray}64 PE \\
 & Scheduler  & 0.24 & 0.07 & -- \\
 & \cellcolor{blendgray}SRAM
   & \cellcolor{blendgray}0.43
   & \cellcolor{blendgray}0.06
   & \cellcolor{blendgray}91 KB \\
\cmidrule(lr){2-5}
\vspace{0.25ex}
 & Total      & 4.44 & 1.59 & -- \\
\midrule
\multirow{4}{*}{GCC}
 & P \& S     & 0.72 & 0.30 & -- \\
 & \cellcolor{blendgray}Blending
   & \cellcolor{blendgray}0.96
   & \cellcolor{blendgray}0.44
   & \cellcolor{blendgray}64 PE \\
 & \cellcolor{blendgray}SRAM
   & \cellcolor{blendgray}1.01
   & \cellcolor{blendgray}0.05
   & \cellcolor{blendgray}190 KB \\
\cmidrule(lr){2-5}
\vspace{0.25ex}
 & Total      & 2.69 & 0.79 & -- \\
\bottomrule
\end{tabular}%
}
\end{minipage}
\hfill
\begin{minipage}[t]{0.49\columnwidth}
\centering
\textbf{(b) DeGS}\par\smallskip
\resizebox{\linewidth}{!}{%
\begin{tabular}{l c c c}
\toprule
 
Component & Area & Power & Config \\
\midrule
Projection Unit   & 0.92 & 0.60  & 4 \\
Sorting Unit      & 0.02 & 0.01  & 4 \\
\rowcolor{blendgray}
Span Engine       & 0.17 & 0.09  & 2+8 \\
\rowcolor{blendgray}
Packing Scheduler & 0.07 & 0.04  & 8 \\
\rowcolor{blendgray}
Blending Array    & 0.88 & 0.23  & 64 PE \\
\cmidrule(lr){1-4}
 
Compute Units     & 2.06 & 0.97  & -- \\
\midrule
Gaussian Buffer   & 0.31 & 0.01  & 72 KB \\
Feature Buffer    & 0.03 & 0.001 & 8 KB \\
\rowcolor{blendgray}
Span Buffer       & 0.04 & 0.001 & 2 KB \\
\rowcolor{blendgray}
Packing Buffer    & 0.16 & 0.02  & 32 KB \\
\rowcolor{blendgray}
Pixel Buffer      & 0.12 & 0.01  & 16 KB \\
\cmidrule(lr){1-4}
 
SRAM              & 0.66 & 0.04  & 130 KB \\
\midrule
 
Total             & 2.72 & 1.01  & -- \\
\bottomrule
\end{tabular}%
}

\end{minipage}
\raggedright\small
\colorbox{blendgray}{\textit{Note:}} 
Shaded entries indicate the components whose area and power increase when scaling the compute resources of the blending stage.
\end{table}

\subsection{Methodology}

\textbf{Workloads \& Datasets.} To ensure fair comparison with our baselines, we adopt standard benchmarks commonly used in prior 3DGS accelerator works~\cite{gscore, gbu, gcc, nebula, metasapiens, orange, gsnorm, gaurast} and select representative scenes from Tanks \& Temples~\cite{knapitsch2017tanks}, Deep Blending~\cite{hedman2018deep}, and Mip-NeRF 360~\cite{barron2022mip}. These datasets cover dense and sparse Gaussian distributions across both indoor and outdoor scenes. All scenes are trained for 30K iterations using the original 3DGS pipeline~\cite{3dgs}. We extended the output resolution from 720p to 8K, ensuring the test covers the effects of resolution growth on array utilization, bandwidth pressure, and system throughput, while also ensuring comparability with existing evaluations.

\textbf{Baselines.} We compare DeGS against representative domain-specific 3DGS rendering accelerators in terms of throughput, energy efficiency, and scalability. GSCore~\cite{gscore} is the earliest dedicated 3DGS accelerator and largely follows a GPU-style tile-based dataflow. Although GBU~\cite{gbu} was originally proposed as a GPU-coprocessor, we include it because of its distinctive IRSS dataflow. For fairness, we compare against the standalone GBU architecture reported in the original paper, and all later references to GBU in this section follow this standalone configuration. GCC~\cite{gcc} is a specialized architecture based on Gaussian-wise and cross-stage dataflow reorganization. Table~\ref{tab:t1} further compares the area and power breakdowns of these designs. Across all architectures, area is mainly dominated by the projection and sorting (P\&S) units, the Gaussian blending unit (Blending), and on-chip SRAM. The original GBU configuration uses only 8 blending PEs, which leads to high utilization but also makes the blending unit a relatively small fraction of total area, resulting in low throughput and high latency regardless of whether area normalization is applied in our experiment. To enable a fairer comparison, we scale GBU to 64 PEs, matching the other baselines.  We ensure that DeGS remains comparable to these baselines in both area and power. 

\textbf{Simulation.} We build cycle-accurate simulators for DeGS and all baseline architectures, modeling key timing behaviors including dataflow scheduling, pipeline bubbles, and SRAM access conflicts to obtain accurate throughput, latency and effective utilization. 
\textcolor{black}{The simulators are trace-driven: all architectures consume the same projected and depth-ordered Gaussian workloads, including Gaussian--tile associations, candidate-fragment coverage, per-tile depth order, and per-pixel termination behavior. They advance execution cycle by cycle and model pipeline backpressure, issue scheduling, conflict delays, PE execution, pixel-state read/writeback, off-chip bandwidth limits, and SRAM capacity and port/bank conflicts.}
For image quality evaluation, we use the rendering output of the original 3DGS implementation running on an RTX 3090~\cite{nvidia3090} as the reference. For fair comparison, we align the computational target frequency, on-chip storage budget, and process parameters across all architectures. Unless otherwise specified, we use an LPDDR5~\cite{lpddr5} memory configuration (6400 MT/s, 256-bit bus width), comparable to NVIDIA Orin AGX~\cite{orin}. In sensitivity analysis, we additionally evaluate a lower-bandwidth configuration (e.g., 3200 MT/s, 128-bit), which is closer to those used in prior work. 

\begin{figure}[t]
  \centering
  \includegraphics[width=0.48\textwidth]{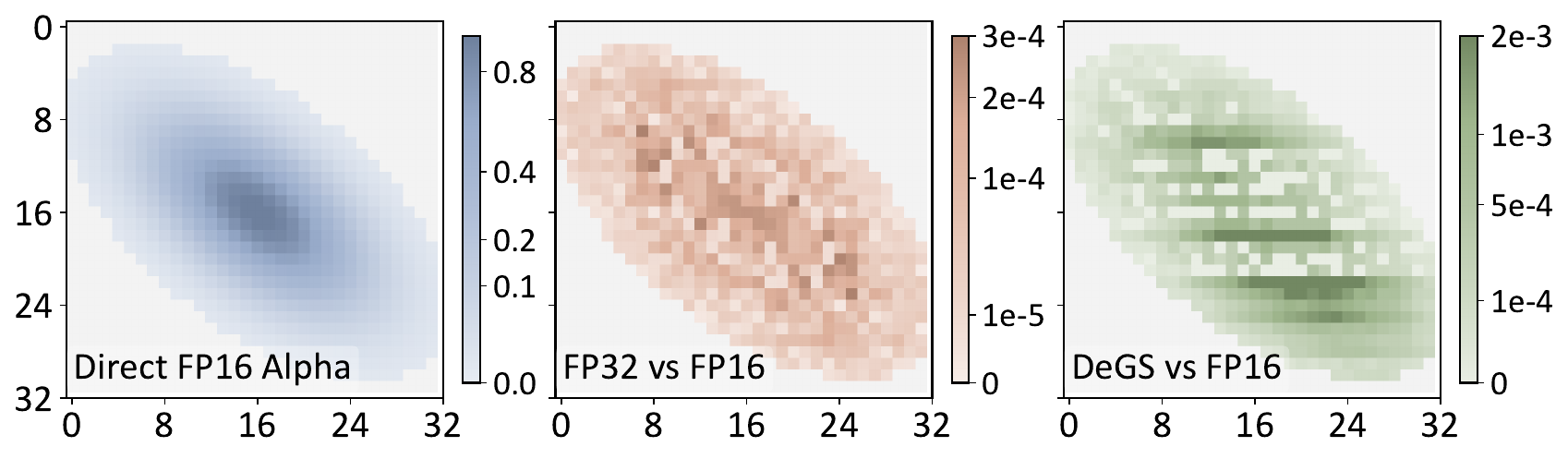}
  \caption{Error of Single-Gaussian Parsing.}
  \label{fig:fig7}
\end{figure}

\begin{figure}[t]
  \centering
  \includegraphics[width=0.48\textwidth]{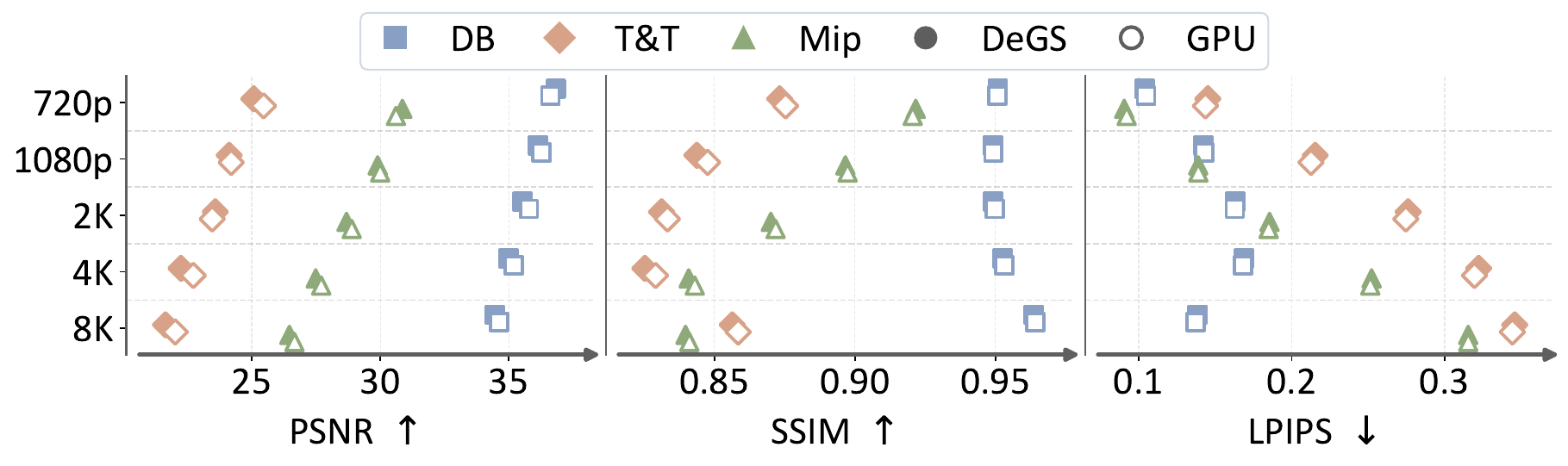}
  \caption{End-to-End Rendering Quality.}
  \label{fig:fig8}
\end{figure}

\textbf{Implementation.} We implement the core logic of DeGS in SystemVerilog at the RTL level and synthesize the design using Synopsys Design Compiler under the TSMC 28\,nm standard-cell library. To align with all baselines, the target clock frequency is set to 1.0\,GHz. On-chip storage structures are modeled using CACTI-P~\cite{cacti-p} under the same 28\,nm technology to obtain area, leakage power, and dynamic read/write energy. All hardware comparisons are conducted under aligned frequency, technology node, and on-chip storage budget.


\subsection{Rendering Quality}

\textbf{Numerical Error of Single Gaussian Parsing.}
Unlike methods such as pruning, compression, or frame reuse that improve efficiency by changing the rendered result, DeGS preserves the mathematical semantics of standard 3DGS blending. Therefore, any observed deviation arises from limited numerical precision and LUT implementation. 
To isolate this numerical effect, Figure~\ref{fig:fig7} evaluates the $\alpha$ map of a single projected $32\times32$ Gaussian with Eq.~\ref{eq:6}. 
Relative to direct FP32 evaluation, direct FP16 computation yields an average absolute $\alpha$ error of $6.75\times10^{-5}$. When the DeGS is further applied as Eq.~\ref{eq:18}, the average absolute error relative to direct FP16 becomes $2.72\times10^{-4}$.
These results show that DeGS preserves the per-Gaussian $\alpha$ distribution, with only minor perturbations near the support boundary. At high resolutions, the main numerical challenge comes not from the semantics-preserving DDA reformulation, but from the enlarged dynamic range of screen-space coordinates and local increments in boundary-related computation. DeGS therefore applies lightweight mixed-precision support on these critical paths, and represents large-magnitude intermediate values in fixed point to improve numerical robustness. As a result, boundary and offset computation remains stable at 4K--8K without introducing high precision overhead to the full pipeline.

\textbf{End-to-End Rendering Quality.}
We next compare the absolute rendering quality of DeGS against the GPU reference. Figure~\ref{fig:fig8} reports PSNR, SSIM, and LPIPS from 720p to 8K on Deep Blending, Tanks\&Temples, and Mip-NeRF 360. Across all tested settings, DeGS closely follows the GPU reference in all three metrics. The average absolute gap / worst-case degradation is 0.225\,dB / 0.478\,dB in PSNR, 0.00155 / 0.00381 in SSIM, and 0.00112 / 0.00273 in LPIPS. DeGS also follows the same resolution-dependent quality trend as the GPU reference across all dataset groups, with no visible systematic drift.
These results indicate that the reformulated execution flow does not introduce measurable end-to-end quality loss, and is unlikely to cause visually noticeable artifacts in the rendered results.
Instead, the remaining differences are consistent with the implementation-level numerical effects observed in the single-Gaussian study. With mixed-precision protection on numerically sensitive screen-space paths, DeGS preserves stable rendering quality from 720p to 8K.

\subsection{Performance}
\label{subsec:performance}

\begin{figure}[t]
  \centering
  \includegraphics[width=0.48\textwidth]{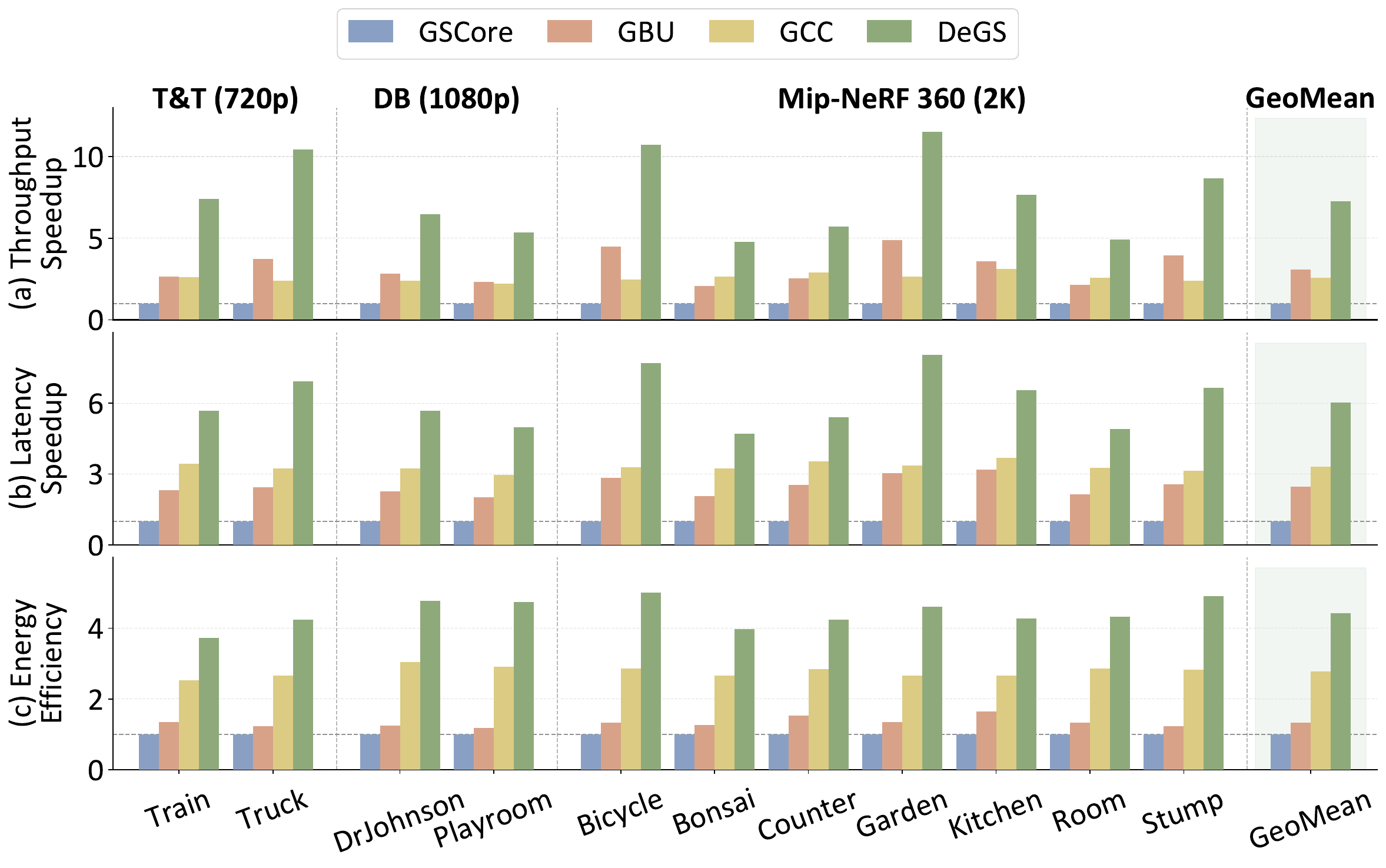}
 \caption{Area-normalized improvement factors across end-to-end latency speedup, steady-state throughput gain, and sustained energy-efficiency gain, respectively. Higher is better in all cases.}
  \label{fig:fig9}
\end{figure}

We evaluate hardware performance using configurations in Table~\ref{tab:t1}. Steady-state throughput measures sustained frame delivery under stage-overlapped execution, and thus reflects whether the architecture can maintain a stable stream of useful work in continuous rendering scenarios. End-to-end latency measures the response time of a single rendering request from input to final image, and therefore captures interactive responsiveness under abrupt viewpoint changes. Sustained energy efficiency further evaluates how effectively the architecture converts silicon area and energy into rendered frames during long-running execution, which is especially important for edge deployment. The gain is most pronounced on large outdoor scenes such as \textit{garden} and \textit{bicycle}, where irregular Gaussian coverage and asynchronous termination are stronger, while the gap becomes smaller on dense indoor scenes such as \textit{bonsai}, \textit{room}, and \textit{playroom}.

As shown in Figure~\ref{fig:fig9}, DeGS consistently outperforms prior architectures. For steady-state throughput, DeGS improves over GBU by 2.14$\times$--2.80$\times$ across scenes, with a geometric-mean gain of 2.36$\times$. Relative to GSCore and GCC, the gains are 7.25$\times$ and 2.82$\times$, respectively.
For end-to-end latency, DeGS achieves 1.45$\times$--2.39$\times$ speedup over GCC across scenes, with geometric-mean gains of 6.02$\times$, 2.44$\times$, and 1.82$\times$ over GSCore, GBU, and GCC, respectively. DeGS removes invalid work before blending and regularizes the remaining tasks before they reach the arithmetic array, so under pipelined execution the blending stage is no longer dominated by spatial and temporal redundancy.
For sustained energy efficiency, DeGS improves over the strongest baseline by 1.47$\times$--1.75$\times$ across scenes, with geometric-mean gains of 4.42$\times$, 3.32$\times$, and 1.59$\times$ over GSCore, GBU, and GCC, respectively. Although DeGS does not introduce a separate frame or Gaussian reuse algorithm, it benefits more effectively from larger tile sizes because a larger aggregation window reduces Gaussian duplication and scheduling overhead, while the decoupled dataflow avoids the extra spatial and temporal redundancy that grows with tile size in prior coupled architectures.

\begin{figure}[t]
  \centering
  \includegraphics[width=0.48\textwidth]{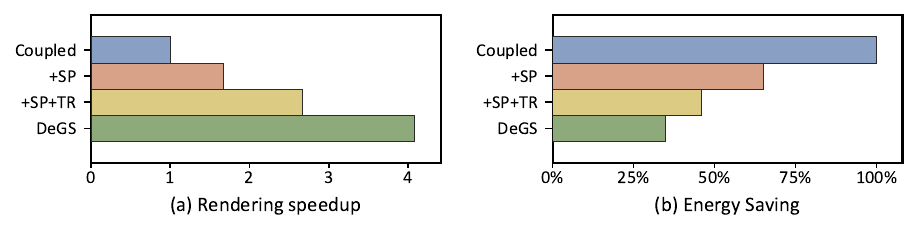}
  \caption{\textcolor{black}{Mechanism-level speed and energy ablation averaged across datasets.}}
  \label{fig:energy_speed_ablation_rebuttal}
\end{figure}

\textcolor{black}{\textbf{Ablation study.} Figure~\ref{fig:energy_speed_ablation_rebuttal} reports a mechanism-level attribution, since the three DeGS stages are connected through Span Vectors, conflict-domain scheduling, and dense-packet interfaces. The latency benefit mainly appears after the full pipeline is enabled. Span Parsing removes spatial redundancy, while Task Reorganization and Blending Execution convert the remaining valid spans into dense backend execution. For energy, the largest reduction comes from Span Parsing, because early removal of invalid fragments avoids downstream $\alpha$ evaluation, pixel-state accesses, and backend activity, even after accounting for frontend and buffer overhead.}

\begin{table}[t]
\centering
\caption{\textcolor{black}{Trace-derived 4DGS workload evidence and DeGS backend benefit at $960\times720$, $16\times16$ tile size, and 64 PEs.}}
\label{tab:dynamic_4dgs}
\setlength{\tabcolsep}{3.2pt}
\renewcommand{\arraystretch}{1.08}
\resizebox{\linewidth}{!}{%
\begin{tabular}{lcccc}
\toprule
\textbf{Metric} & \textbf{Coffee} & \textbf{Spinach} & \textbf{Beef} & \textbf{Mean} \\
\midrule
\multicolumn{5}{l}{\textbf{Dynamic and irregular 4DGS rendering traces}} \\
Active-Gaussian overlap / tile-workload churn
& 0.450 / 4.6\% & 0.270 / 4.9\% & 0.266 / 5.7\% & 0.329 / 5.1\% \\
P99 tile candidates / footprint Gini
& 2.39K / 0.854 & 1.95K / 0.882 & 2.06K / 0.883 & 2.13K / 0.873 \\
\midrule
\multicolumn{5}{l}{\textbf{DeGS rendering-stage benefit}} \\
Blending PE utilization, coupled / DeGS
& 15.5\% / 86.7\% & 27.4\% / 92.3\% & 28.6\% / 91.5\% & 23.9\% / 90.2\% \\
Rendering speedup / trace-level E2E proxy
& 5.94$\times$ / 2.51$\times$ & 3.16$\times$ / 1.95$\times$ & 3.03$\times$ / 1.95$\times$ & 4.04$\times$ / 2.20$\times$ \\
\bottomrule
\end{tabular}}
\end{table}

\textcolor{black}{\textbf{Dynamic Gaussian Splatting applicability.}
We use 4DGS as a representative dynamic-scene workload to test whether DeGS extends beyond static 3DGS rendering. We train N3V scenes with the standard 4DGS pipeline, materialize timestamp-conditioned Gaussian attributes, and replay the resulting frame-local traces. Table~\ref{tab:dynamic_4dgs} shows that these traces preserve the key workload properties assumed in our motivation: active-Gaussian overlap is only 0.266--0.450, tile-workload churn is 4.6\%--5.7\%, and projected coverage remains highly irregular. Since alpha blending is still a frame-local near-to-far recurrence, cross-frame motion changes the trace distribution but does not introduce cross-frame dependence into the Blending Array. DeGS therefore retains its rendering-stage benefit on dynamic scenes, improving mean PE utilization from 23.9\% to 90.2\%, achieving 4.04$\times$ rendering speedup and 2.20$\times$ end-to-end speedup when dynamic materialization cost is held unchanged.}

\begin{figure}[t]
  \centering
  \includegraphics[width=0.48\textwidth]{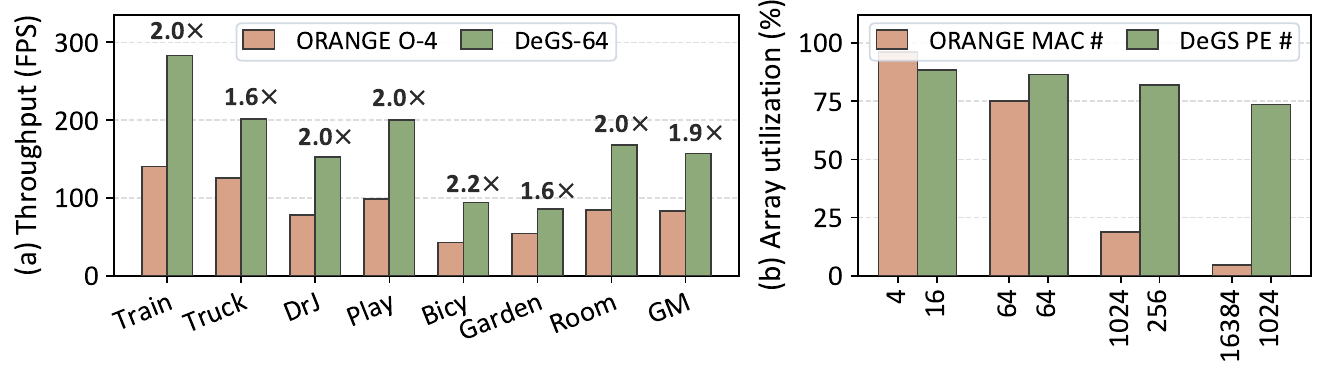}
  \vspace{-5pt}
  \caption{\textcolor{black}{Raw throughput and backend utilization versus ORANGE.}}
  \label{fig:orange_degs_area_util}
\end{figure}

\textcolor{black}{\textbf{Contemporary NPU baseline.}
Beyond the three rasterization-oriented 3DGS accelerator dataflows evaluated above, we further compare DeGS with ORANGE~\cite{orange}, a contemporary NPU-based 3DGS rendering baseline. Figure~\ref{fig:orange_degs_area_util}(a) reports raw end-to-end throughput on the same seven scenes without area normalization; DeGS-64 still outperforms ORANGE O-4 by 1.6$\times$--2.2$\times$ while using a substantially smaller area footprint. Figure~\ref{fig:orange_degs_area_util}(b) shows that widening a GEMM-style backend alone is insufficient: ORANGE reformulates blending into a GEMM-friendly kernel with a narrow reduction dimension ($K=6$), so scaling the systolic backend from 4 to 16,384 MACs reduces local array utilization from 96.0\% to 4.69\%. In contrast, DeGS regularizes valid blending work before execution and maintains 88.5\%--73.6\% mean backend utilization when scaling from 16 to 1024 PEs. This confirms that DeGS addresses a complementary dataflow bottleneck rather than relying on a wider general-purpose NPU array to absorb 3DGS irregularity.}

\label{sec:scaling}

\begin{table}[t]
\centering
\caption{Coordinated scaling configurations of DeGS.}
\label{tab:t2}
\setlength{\tabcolsep}{4pt}
\renewcommand{\arraystretch}{1.08}
\resizebox{\linewidth}{!}{%
\begin{tabular}{c c c c c c c}
\toprule
Total PEs & Peak TOPS & Tile Size & \#GSUs & \#Span Units & \#Issue Banks & PEs/Array \\
\midrule
16   & 0.1 & 16  & 1 & 4  & 4  & 4  \\
64   & 0.5 & 32  & 2 & 8  & 8  & 8  \\
256  & 2.0 & 64  & 4 & 16 & 16 & 16 \\
1024 & 8.0 & 128 & 8 & 32 & 32 & 32 \\
\bottomrule
\end{tabular}}
\end{table}

\subsection{Scalability Analysis}
Scaling in DeGS is not achieved by simply replicating a wider blending PE array. The key architectural advantage is that DeGS removes invalid work before blending execution and localizes dependence handling in the scheduling layer, so increasing array scale no longer amplifies spatial redundancy and temporal redundancy in tandem, as in prior coupled designs.

\textbf{Decoupled Scaling Rule.}
This property of DeGS enables the coordinated scaling rule in Table~\ref{tab:t2}. The rationale is that Projection and Sorting are primarily constrained by off-chip bandwidth and therefore are not the main scaling bottlenecks. The Span Parsing and Task Reorganization stages only need to expose valid work in row-level parallelism. Moreover, as Figure~\ref{fig:fig1}(a) suggests, a projected Gaussian typically expands into multiple valid rows, and each valid row usually contains multiple effective pixels rather than isolated single-pixel work. DeGS therefore provisions the numbers of GSUs, Span Units, and Issue Banks for stable packet generation instead of one-to-one matching with the PE count, while leaving moderate redundancy in these stages to absorb workload imbalance. Consequently, these stages can be provisioned sublinearly with PE count in our evaluated design space. By contrast, the Pixel Buffer must scale linearly with PE count. The effectiveness of this asymmetric provisioning is reflected in Figure~\ref{fig:fig11}: across most coordinated scaling points, the pressure of GSU, Span Array, and Issue Bank remains below full starvation, and they become limiting mainly in the low-resolution, large-array regime.

\begin{figure}[t]
  \centering
  \includegraphics[width=0.48\textwidth]{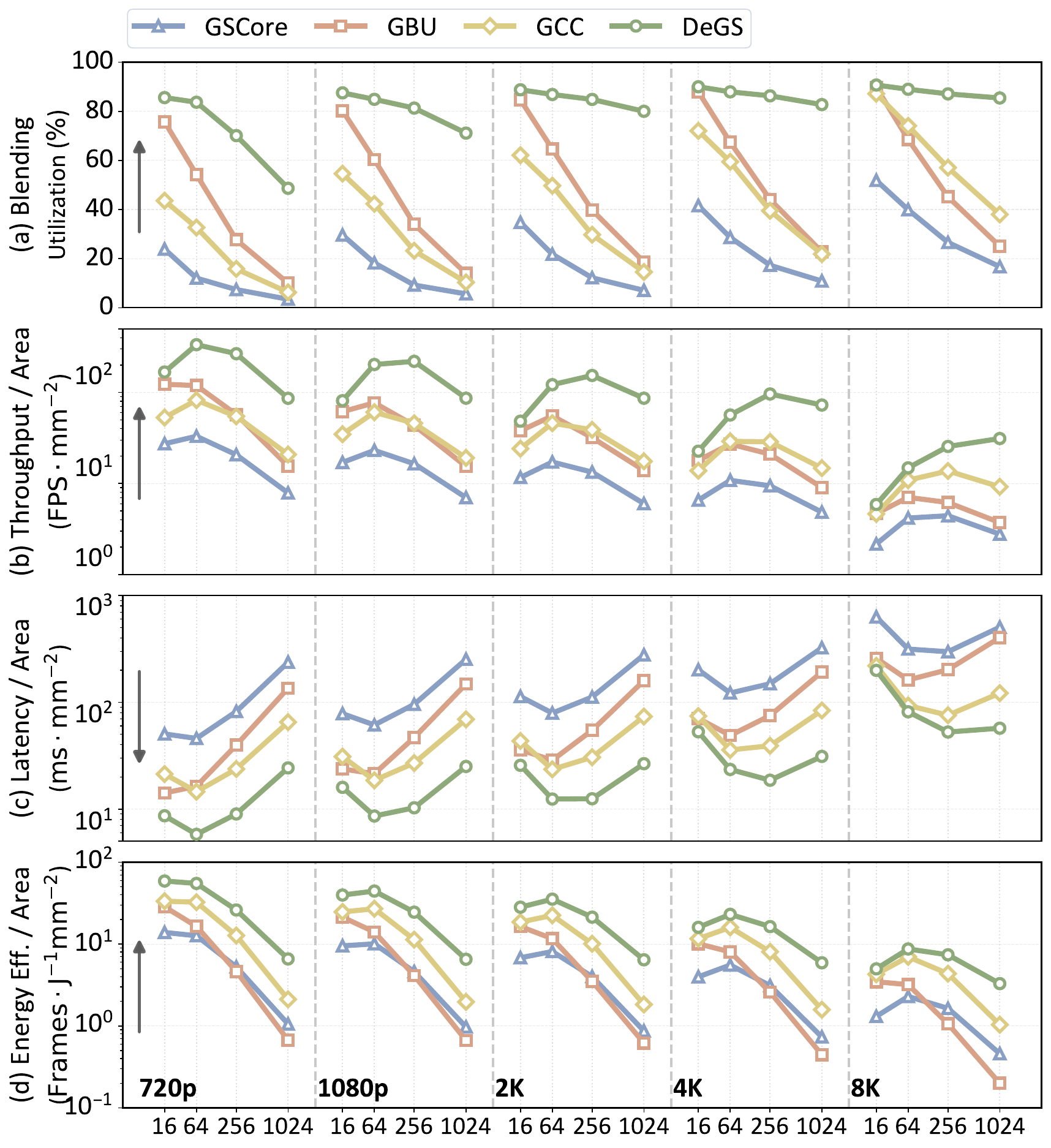}
  \caption{Blending PE utilization of each accelerator.}
  \label{fig:fig10}
\end{figure}

\begin{figure}[t]
  \centering
  \includegraphics[width=0.48\textwidth]{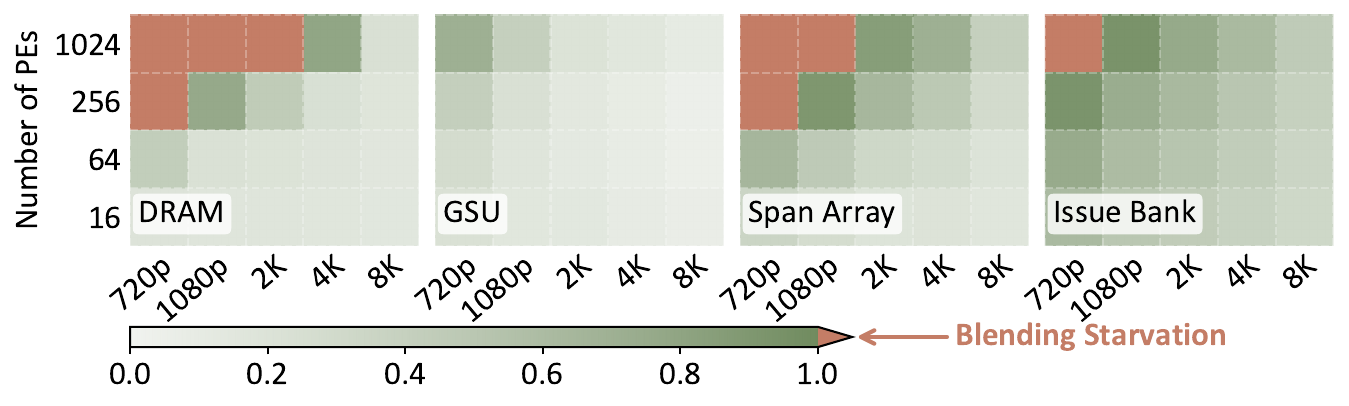}
  \caption{Bottleneck pressure of pre-execution stages.}
  \label{fig:fig11}
\end{figure}

\textbf{Scaling Behavior.}
From Figure~\ref{fig:fig10}, DeGS not only delivers better metrics across almost all evaluated resolutions and array sizes, but more importantly degrades much more gracefully as the array scales up and enters the effective scaling regime at a lower resolution than prior designs. In particular, DeGS maintains over 80\% PE utilization at higher resolutions, whereas the utilization of other architectures quickly drops to 10\%--30\% as the array becomes larger. Correspondingly, GSCore, GCC, and GBU all begin to show degradation in area-normalized throughput, latency, or energy efficiency at relatively small array scales. These results indicate that the decoupled dataflow of DeGS cuts off the propagation path through which wider arrays and larger blocks amplify redundancy in prior architectures.

\textbf{Bottleneck Analysis.}
Figure~\ref{fig:fig11} further shows that, when the array becomes large while the rendering resolution remains low, the saturation of DeGS is mainly caused by insufficient workload supply of valid tasks, rather than by backend redundancy. In this regime, each Gaussian produces shorter valid row spans, causing the Span Engine to reach its supply limit more easily. At the same time, the number of scheduler-visible tasks in the Issue Banks also decreases, which makes local scheduling holes and backend waiting more likely, and therefore prevents the Blending Array from being continuously saturated. However, once the array throughput exceeds the rate at which off-chip DRAM can provide valid data to the blending stage, simply increasing the number of frontend units can no longer deliver proportional benefit under such workloads.

\subsection{Sensitivity Study}

\begin{figure}[t]
  \centering
  \includegraphics[width=0.48\textwidth]{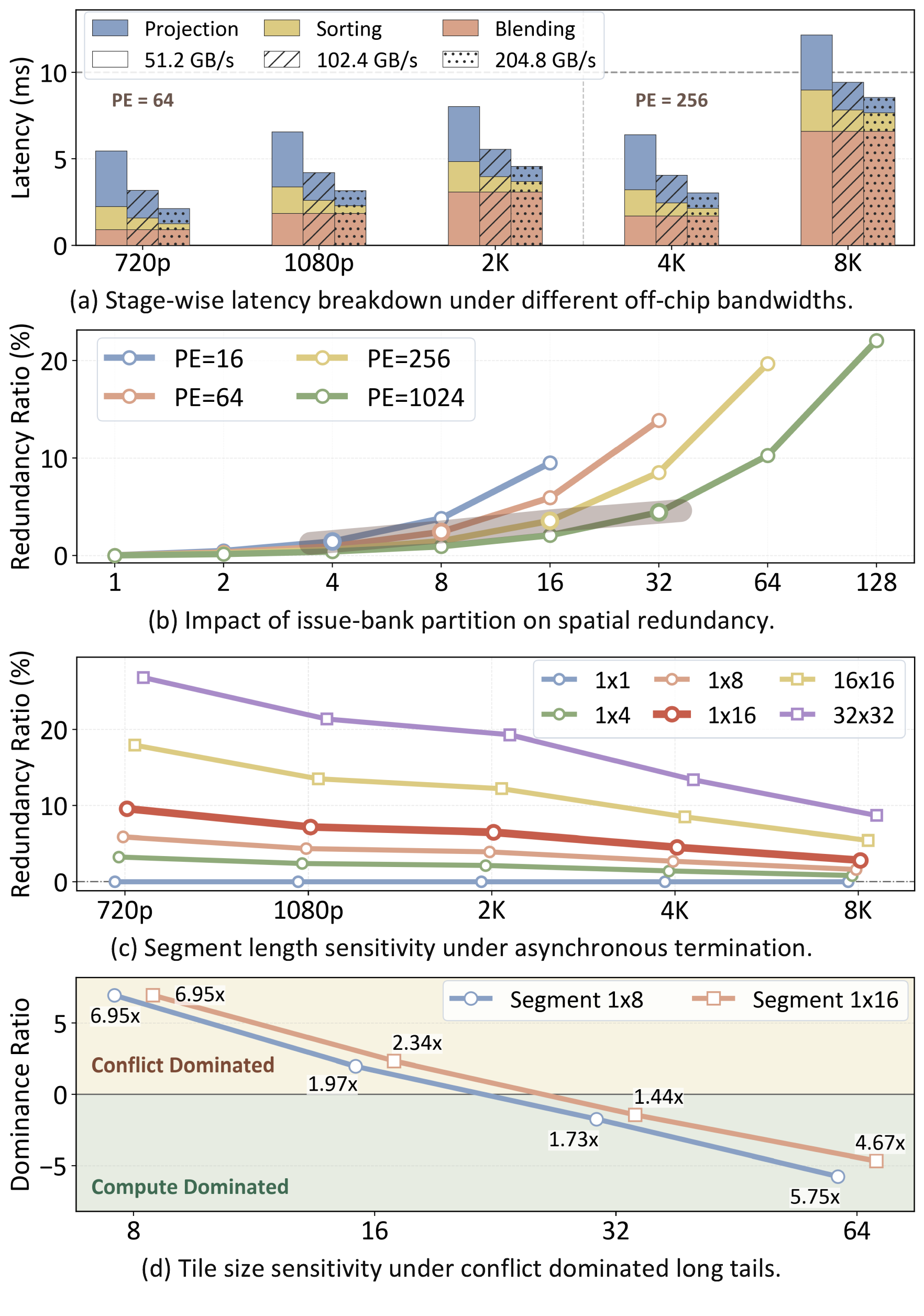}
  \caption{Sensitivity Study.}
  \label{fig:fig12}
\end{figure}

\textbf{Off-chip Bandwidth Sensitivity.} Figure~\ref{fig:fig12}(a) shows the stage-wise latency trend of DeGS under three off-chip bandwidth settings at two representative array scales (PE=64 and PE=256) on Mip-NeRF 360. Increasing bandwidth from 51.2\,GB/s to 204.8\,GB/s consistently reduces the latency of Projection and Sorting across all resolutions. As a result, the system bottleneck progressively shifts away from preprocessing and toward rendering as bandwidth increases. This trend is most evident in the high-load regime: for the 8K workload at PE=256, Blending remains the dominant stage even at 204.8\,GB/s. \textcolor{black}{The 100 Hz real-time reference line further shows that DeGS achieves real-time rendering for resolutions up to 4K under both low- and high-bandwidth settings.}
These results indicate that lower-bandwidth settings would make DeGS appear memory-bound and thus obscure the scaling behavior of the blending backend. Therefore, our main evaluation uses an Orin-class LPDDR5 configuration to place the system in the regime where scalability is determined primarily by blending dataflow organization rather than by DRAM bandwidth. 

\textbf{Bank Partition Sensitivity.} Figure~\ref{fig:fig12}(b) shows the spatial redundancy ratio at 1080p as the number of banks increases under different PE configurations on Mip-NeRF 360. For all PE configurations, redundancy increases as the number of banks grows, and the growth is much faster at larger PE counts. 
This trend indicates that finer bank partitioning simplifies localized dispatch and storage management, but also fragments spatially correlated Gaussian work and distributes effective tasks less evenly across banks. This uneven distribution lowers the effective task supply seen by blending and ultimately increases spatial redundancy. By contrast, the coordinated scaling points used in DeGS, namely 4/8/16/32 banks for PE=16/64/256/1024, keep the redundancy within roughly 5\%. 

\textbf{Segment Length Sensitivity.} Figure~\ref{fig:fig12}(c) reports the redundancy ratio under different segment length across resolutions with 64 PE setting on Mip-NeRF 360. Finer segments consistently reduce redundancy, with the largest gains appearing at lower resolutions. At higher resolutions, the gap between neighboring settings becomes smaller. This result indicates that segment granularity determines how much termination-induced waiting remains visible to Task Reorganization. Coarse segments expose more temporal imbalance to the scheduler, whereas excessively fine segments provide diminishing redundancy reduction while increasing control overhead. In DeGS, segment sizes on $1\times16$ are sufficient to keep redundancy low, making them a practical operating range.

\textbf{Tile Size Sensitivity.}
Figure~\ref{fig:fig12}(d) reports the signed dominance ratio for the 1080p workload under different tile sizes on Mip-NeRF 360. For both segment settings, small tiles (8 and 16) remain in the upper region, showing that Task Reorganization cannot expose enough conflict-free workload within such a limited scheduling window. As a result, the Blending PE is still dominated by conflict-induced waiting rather than by pure blending throughput. When the tile size increases to 32, the dominance ratio moves close to zero and begins to turn lower region, indicating that the scheduler can now supply a sufficiently dense stream of ready tasks and the system is no longer primarily limited by conflict exposure. Unlike prior coupled architectures, DeGS can scale tile size without incurring substantial spatial or temporal redundancy. The main overhead is only a modest increase in on-chip buffering, which further improves its adaptability to diverse application scenarios.

\section{Related Work}
\label{sec:related_works}

\textbf{Architectural support for 3DGS.}
The rapid adoption of 3DGS has motivated a growing body of hardware and system support for both rendering and training~\cite{gscore, gbu, gcc, vr-pipe, metasapiens, lumina, neo, nebula, hypergs, gauspu, splatonic, durvasula2025arc, gsarch,axis, gsnorm, orange}. On the inference side, prior work has explored tile-based execution~\cite{gscore}, \textcolor{black}{whose rasterization primitive is adopted or extended by many 3DGS accelerators~\cite{gsarch, metasapiens, gsnorm, nebula}}. Row-oriented rasterization and blending support~\cite{gbu}, \textcolor{black}{which reduces coarse spatial waste but remains limited by row-local dependencies and imbalance}. Gaussian-wise execution~\cite{gcc}, \textcolor{black}{which reorganizes cross-stage execution but still couples filtering with backend traversal}. Other works optimize GPU-oriented pipelines~\cite{vr-pipe, gaurast}, caching and reuse~\cite{lumina, neo}, or immersive and large-scale deployment~\cite{hypergs}. \textcolor{black}{In contrast, DeGS decouples valid-work discovery and dependence handling from blending.} A separate set of works targets 3DGS training or SLAM-oriented workloads~\cite{durvasula2025arc, gsarch, gauspu,rtgs, splatonic}, for example by reducing atomic overheads, filtering redundant gradient updates, or exploiting sparsity in online pipelines.

\nocite{cheng2025area,han2025vectee,zheng2025inputsnatch,guan2025mvbsd,xia2025comet}

\textbf{GPU-oriented optimization.}
Another line of work improves 3DGS efficiency by directly reducing the amount of work that needs to be processed on GPU. Representative examples include pruning, compression, quantization, and quality-aware simplification, which remove low-impact Gaussians, reduce memory and storage cost, or trade a controlled amount of rendering fidelity for lower computation and bandwidth demand~\cite{lightgaussian:nips:25, compgs:eccv:24, fang2024mini, eagles:eccv:24, papantonakis:i3d:24}. A separate direction focuses on accelerating Gaussian rendering on commodity GPUs, where optimizations must operate within the constraints of SIMT execution, fixed memory hierarchies, and existing graphics or compute APIs~\cite{vr-pipe, lumina, potamoi, cicero:isca:24, zhao2020deja, zhu2018euphrates, hwang2022cova, hwang2025dejavu, balanced, flashgs, adr, speedy-splat,gs-scale}. These methods typically emphasize scheduling, caching, warping, or temporal reuse on programmable platforms. In contrast to semantics-preserving architectural support for dedicated accelerators, the former reduces the workload itself, while the latter improves how that workload is executed on general-purpose hardware.  

\section{Conclusion}
In this work, we show that the scaling wall of 3DGS rendering is rooted not in insufficient arithmetic throughput alone, but in the coupled execution of $\alpha$-checking and blending, which amplifies irregular spatial coverage and asynchronous termination into severe hardware utilization loss under wider arrays. To address this problem, we propose DeGS, a fully decoupled 3DGS rendering microarchitecture that front-loads valid-work discovery and task regularization through Span Parsing and Task Reorganization, so that backend blending operates only on dense and conflict-free execution packets. Across diverse scenes, resolutions, and array sizes, DeGS consistently delivers higher throughput, energy efficiency, and scaling efficiency than prior semantics-preserving 3DGS accelerators while preserving near-reference rendering quality, suggesting that scalable Gaussian rendering requires redesigning the dataflow before widening blending PEs.

\bibliographystyle{IEEEtran}
\bibliography{sample-base}

\end{document}